\documentclass{appolb}
\usepackage{graphicx}
\usepackage{amsmath}
\usepackage{hyperref}

\begin{document}
\title{A non-perturbative study of the correlation functions of three-dimensional Yang-Mills theory
\thanks{Presented at Excited QCD 2016, 6-12 March 2016, Costa da Caparica, Portugal}%
}
\author{Markus Q. Huber
\address{Institute of Physics, University of Graz, NAWI Graz, Universit\"atsplatz 5, 8010 Graz, Austria}
}

\maketitle

\begin{abstract}
Yang-Mills theory is studied in three dimensions using the equations of motion of the $1$PI and $3$PI effective actions. The employed self-contained truncation includes the propagators, the three-point functions and the four-gluon vertex dynamically. In the gluon propagator also two-loop diagrams are taken into account. The higher gluonic correlation functions show sizable deviations from the tree-level only at low momenta. Also the couplings derived from the vertices agree well down to a few GeV. In addition, different methods to subtract spurious divergences are explored.
\end{abstract}

  
\section{Introduction}

The property of asymptotic freedom, which justifies the use of perturbation theory at high momentum transfer, was crucial to establish QCD as the theory of the strong interaction. However, for many interesting aspects of QCD, like dynamical mass generation, confinement or the description of transitions between the phases of strongly interacting matter, nonperturbative methods are required. One approach among many is functional equations. In their pure form they provide a complete description of QCD, but for concrete calculations they must be truncated. The challenge is then to understand the effect of any truncation and, if necessary, to improve it. 

Understanding the effect of truncations consists in assessing the influence of discarded diagrams and the effects of employed models. Especially the latter is difficult as models are often tailored to produce good results for the calculated quantities. Replacing the models by dynamically calculated correlation functions might thus not lead to a quantitative improvement. For an example, see Ref.~\cite{Blum:2014gna} where the role of the three-gluon vertex in the gluon propagator is discussed. Nevertheless, it is necessary to increase the number of dynamically included quantities and test if the changes become smaller beyond some level in order to obtain a self-consistent and self-contained solution. The recent advances in calculating higher correlation functions and solving systems with dynamical vertices show that this has indeed become feasible by now \cite{Blum:2014gna,Huber:2012kd,Eichmann:2014xya,Binosi:2014kka,Cyrol:2014kca,Huber:2016tvc,Cyrol:2016tym}.

Here I describe the solution of the system of primitively divergent correlation functions of three-dimensional Yang-Mills theory and test the effects of the truncation by varying some of its specifics. More details can be found in Ref.°\cite{Huber:2016tvc}. The advantage over four dimensions is that the theory is UV finite. However, because of the employed cutoff regularization spurious divergences occur in the gluon propagator DSE. Their treatment is alleviated in three dimensions, because the dressing functions fall off polynomially in the UV and the divergences are purely logarithmic and linear. This is in contrast to four dimensions, where these divergences are not quadratic beyond one-loop \cite{Huber:2014tva}. This difference allows an easy subtraction in three dimensions via a fit to the cutoff dependence even when vertices are included dynamically or two-loop diagrams are considered \cite{Huber:2016tvc}.

\begin{figure}[bt]
\begin{minipage}{0.48\textwidth}
\begin{center}
\includegraphics[width=0.9\textwidth]{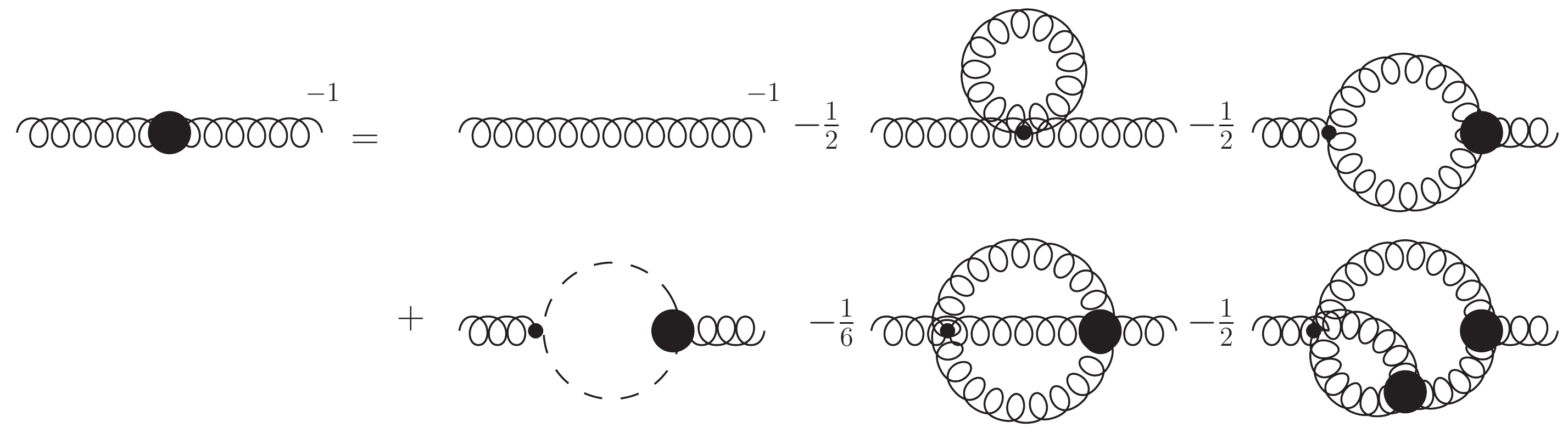}
\end{center}
\end{minipage}
\begin{minipage}{0.48\textwidth}
\begin{center}
\includegraphics[width=0.9\textwidth]{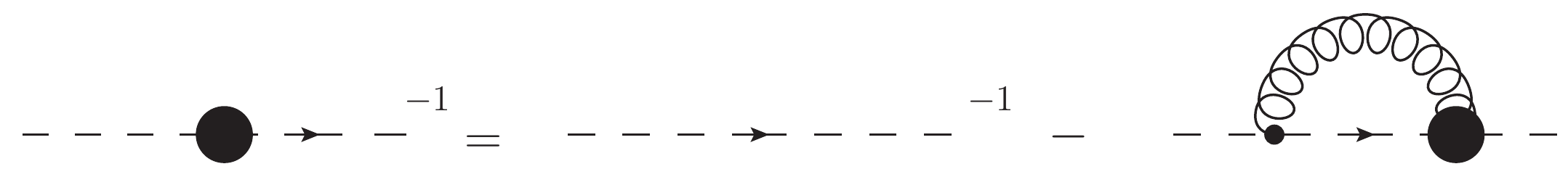}\\
\vskip0mm
\includegraphics[width=0.9\textwidth]{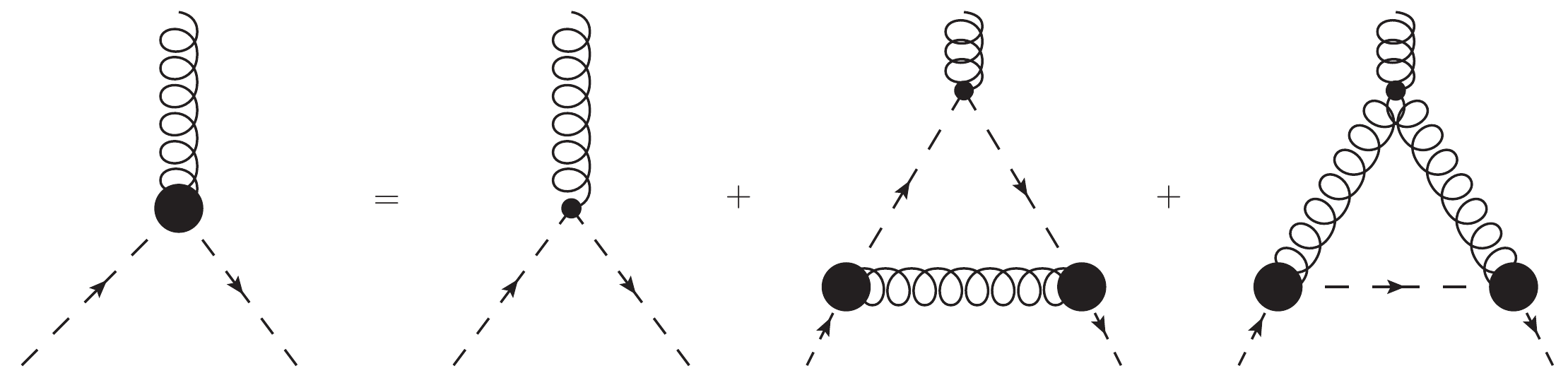}
\end{center}
\end{minipage}
\caption{The full gluon propagator DSE, the full ghost propagator DSE and the truncated ghost-gluon vertex $A$-DSE. Here and in all other figures, internal propagators are dressed, and thick blobs denote dressed vertices, wiggly lines gluons, and dashed ones ghosts.}
\label{fig:gh-gl-ghg}
\end{figure}

\begin{figure}[bt]
\begin{minipage}{0.48\textwidth}
\begin{center}
\vskip2mm
\includegraphics[width=0.9\textwidth]{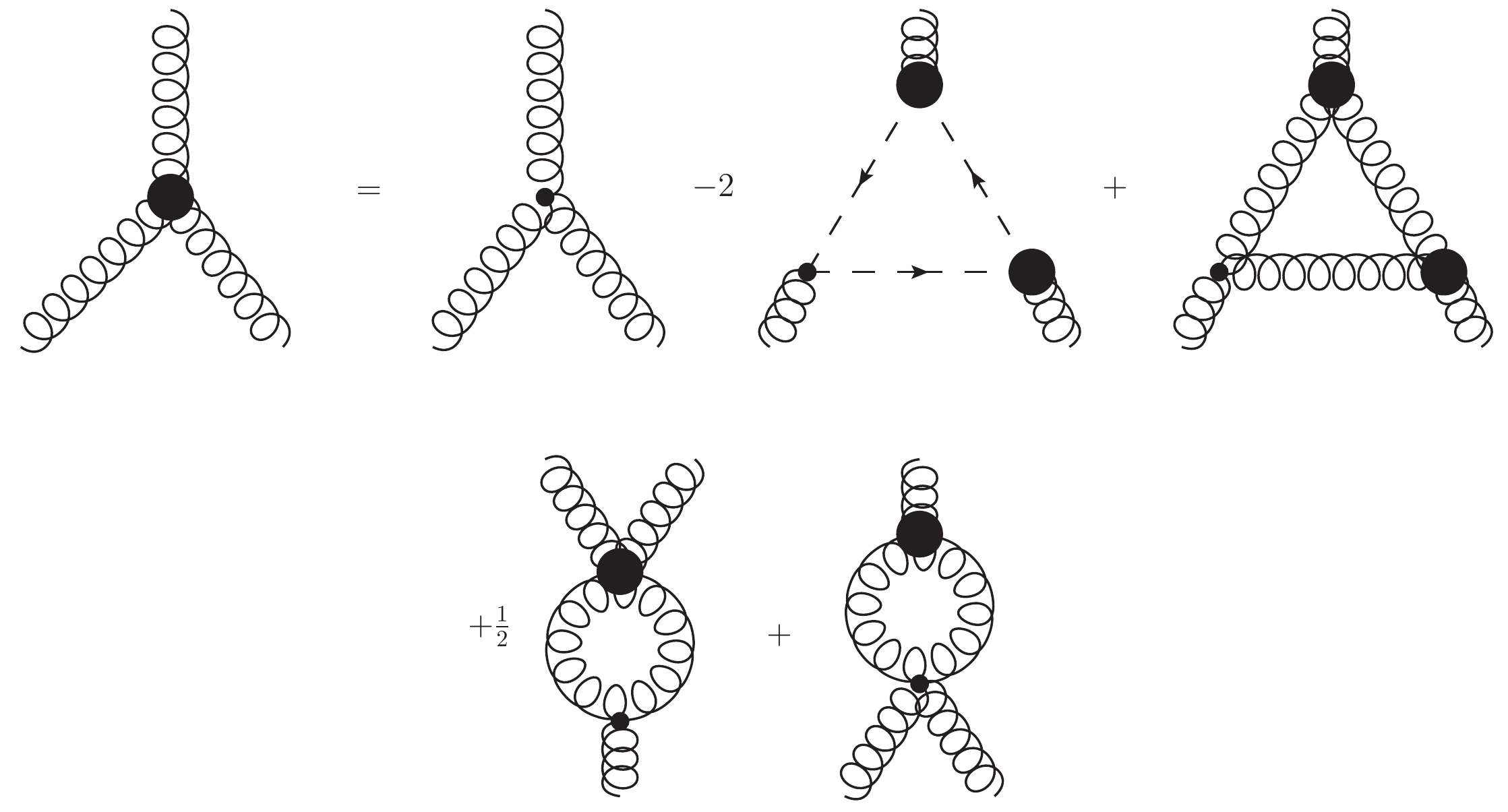}
\end{center}
\end{minipage}
\begin{minipage}{0.547\textwidth}
\begin{center}
\includegraphics[width=0.9\textwidth]{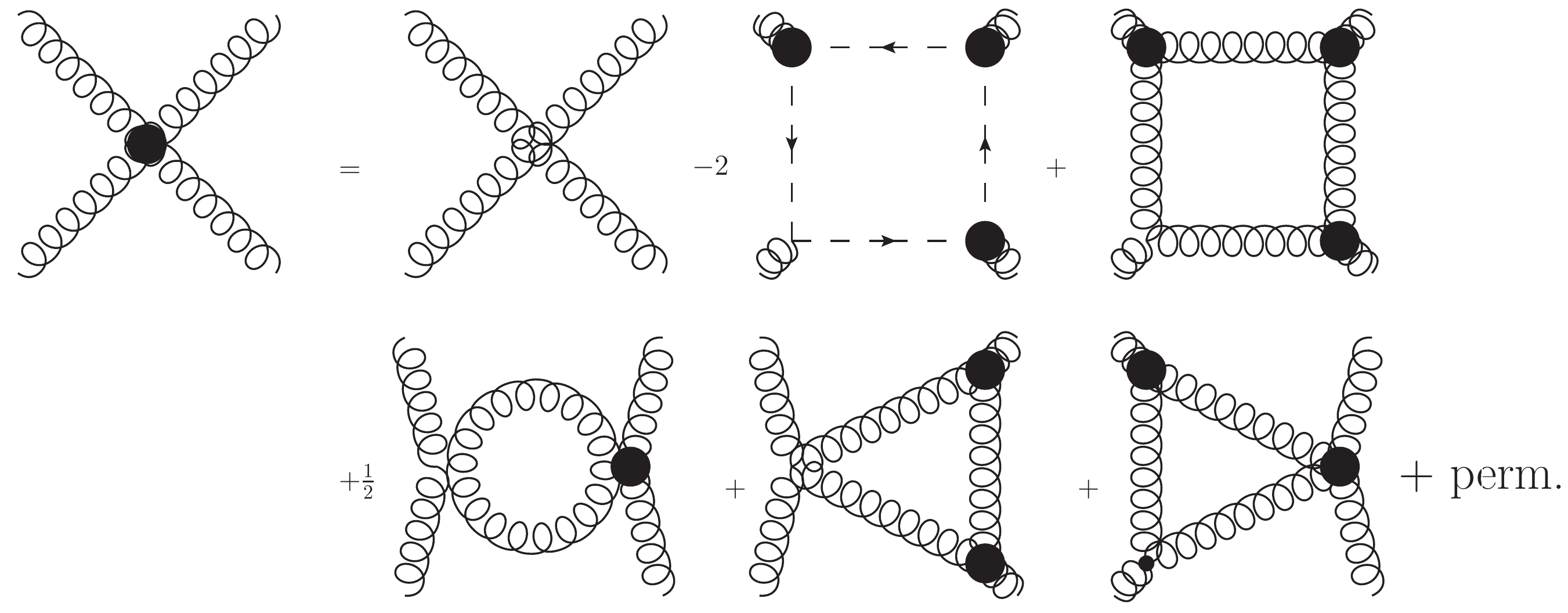}
\end{center}
\end{minipage}
\caption{The truncated three- and four-gluon vertex DSEs.}
\label{fig:3g-4g}
\end{figure}

The truncation considered here includes the propagator, the three-point functions and the four-gluon vertex; see Figs.~\ref{fig:gh-gl-ghg} and \ref{fig:3g-4g} for the corresponding DSEs. Two different sets of equations are employed. One is the equations of motion of the 3PI effective action truncated at three-loops. In this case, only the propagators and the three-point functions are dynamical and the four-gluon vertex is bare. The second set of equations is the Dyson-Schwinger equations (DSEs). Their truncation is specified by setting all non-primitively divergent Green functions to zero. The only freedom left is to choose one of the two DSEs for the ghost-gluon vertex. They are referred to as the $c$- and $A$-DSE, depending on which external leg is attached to the bare vertex. It should be noted that these two equations have in their full form four and twelve diagrams, respectively. However, the specified truncation always reduces them to three diagrams. The truncation is self-contained in the sense that there is no model dependence left and the only parameter is the coupling which sets the scale. 

The equations are solved with a simple fixed point iteration using the framework \textsc{CrasyDSE} \cite{Huber:2011xc} together with \textsc{DoFun} to derive the equations \cite{Alkofer:2008nt,Huber:2011qr}.
To compare with lattice results, the number of colors is set to $2$.

\section{Results}

\begin{figure}[tb]
\centerline{
\includegraphics[width=0.47\textwidth]{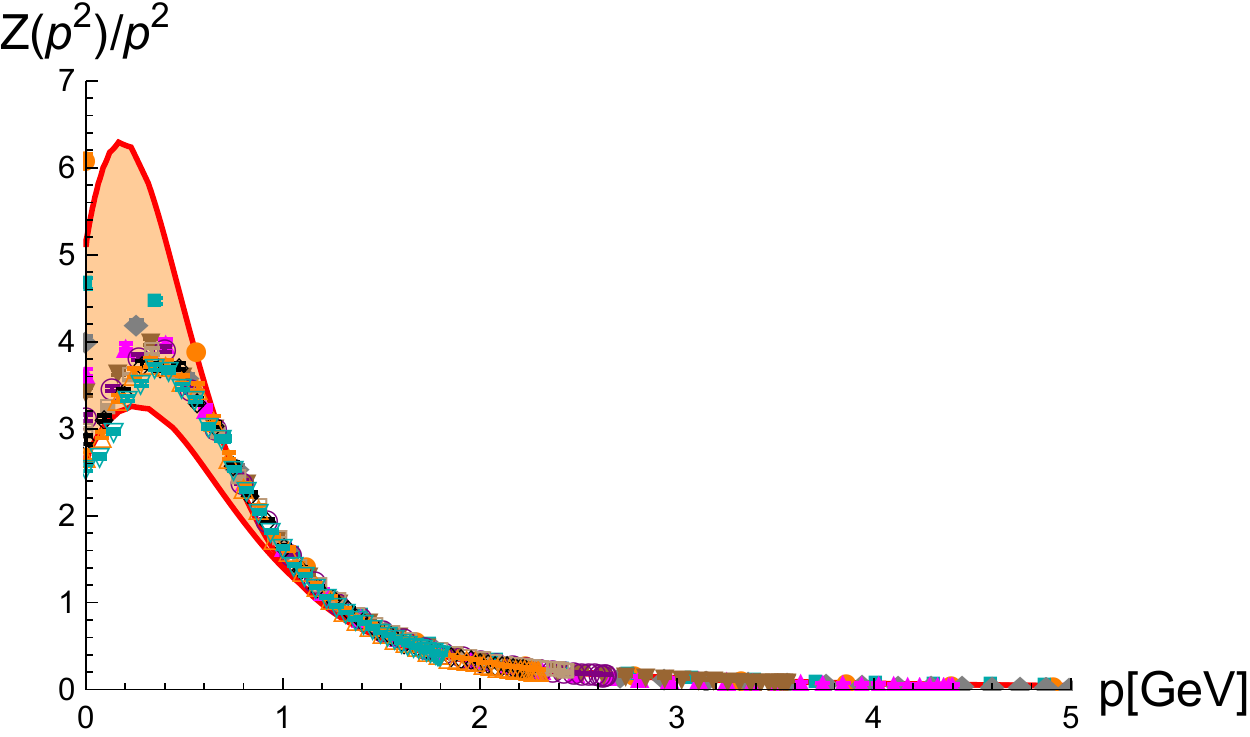}
\hfill
\includegraphics[width=0.47\textwidth]{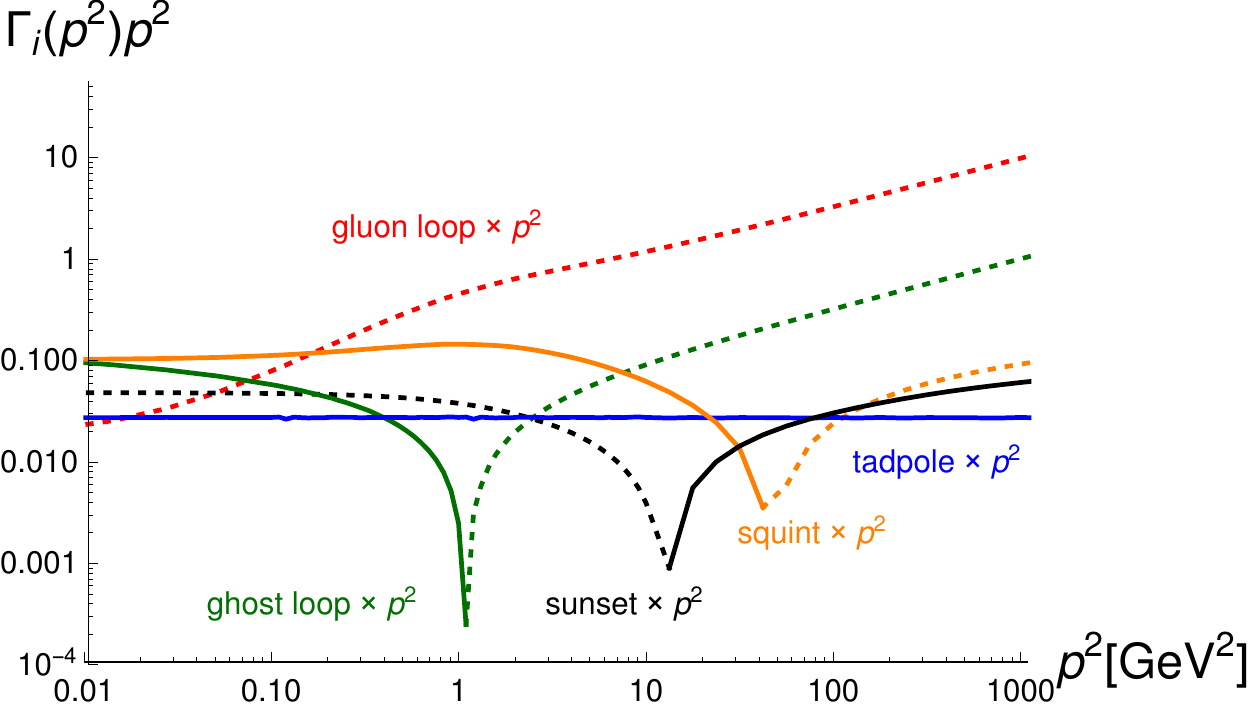}}
\caption{Left: Gluon propagator from the full system in comparison to lattice results \cite{Maas:2014xma}. The band is obtained by varying the maximum of the gluon dressing function between $922$ and $1282\,\text{MeV}$. Right: Contributions of individual diagrams in the gluon propagator DSE.}
\label{fig:res-gl}
\end{figure}

The result for the gluon propagator calculated within the full DSE system is shown in \ref{fig:res-gl}. To relate to a physical scale, which is determined by the dimensionful coupling constant, the gluon dressing function maximum is matched with that of lattice results but taking into account a variation between $90$ and $125\,\%$ because of the deficiency in the region around $1\,\text{GeV}$. A possible reason for this mismatch is the existence of a family of solutions of the equations which is known from four dimensions typically related to a boundary condition for the ghost DSE \cite{Boucaud:2008ji,Fischer:2008uz}. No boundary condition for the ghost is set here as the ghost DSE is finite. 
In Fig.~\ref{fig:res-gl} also the contributions from individual diagrams are shown. As expected, the gluon loop yields the dominant contribution in the UV. Also in the midmomentum regime it clearly dominates. In the IR regime, however, the ghost becomes important. The two-loop diagrams have a clear hierarchy: The sunset diagram contributes only very little, whereas the squint diagram yields the second largest contribution in the midmomentum regime. Indeed, it was found that the squint diagram is important for the stability of the equation under iteration once the gluon bump around $1\,\text{GeV}$ reaches a certain height.

\begin{figure}[tb]
\centerline{
\includegraphics[width=0.47\textwidth]{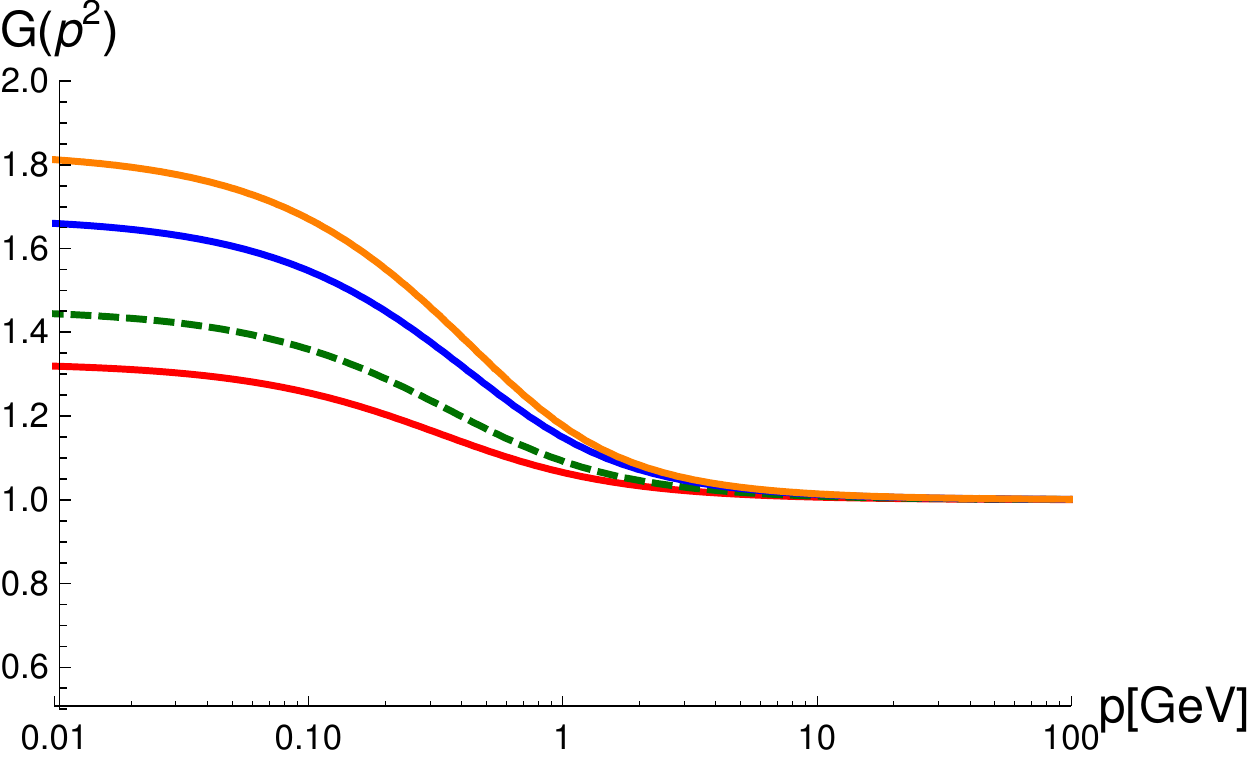}
\hfill
\includegraphics[width=0.47\textwidth]{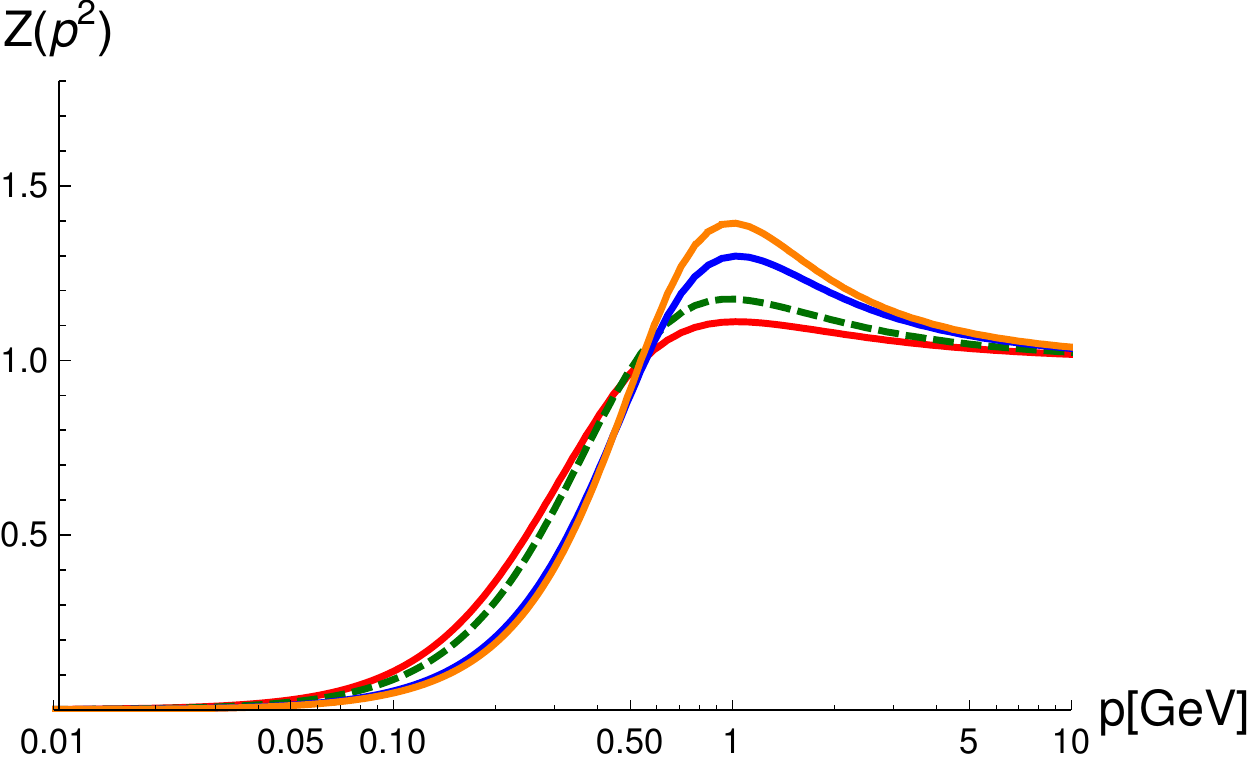}}
\caption{Ghost/Gluon dressing functions from the one-loop truncation using bare ghost-gluon and three-gluon vertices. Different solutions correspond to different values of the gluon propagator at zero momentum.}
\label{fig:res-gh-gl-family}
\end{figure}

To test if different solutions can be obtained, the subtraction of spurious divergences is modified to set the gluon propagator to a specific value at zero momentum. This is motivated by calculations with the FRG where a mass parameter can be varied in the UV leading to different solutions \cite{Cyrol:2016tym}. First results of this method using a one-loop truncation with bare vertices are shown in Fig.~\ref{fig:res-gh-gl-family}. The magnitude of the observed effect has to be taken with a grain of salt, since the model vertices do not vary for different solutions although the full vertices do \cite{Huber:2012kd,Eichmann:2014xya,Cyrol:2014kca,Cyrol:2016tym}.

\begin{figure}[tb]
\centerline{
\includegraphics[width=0.47\textwidth]{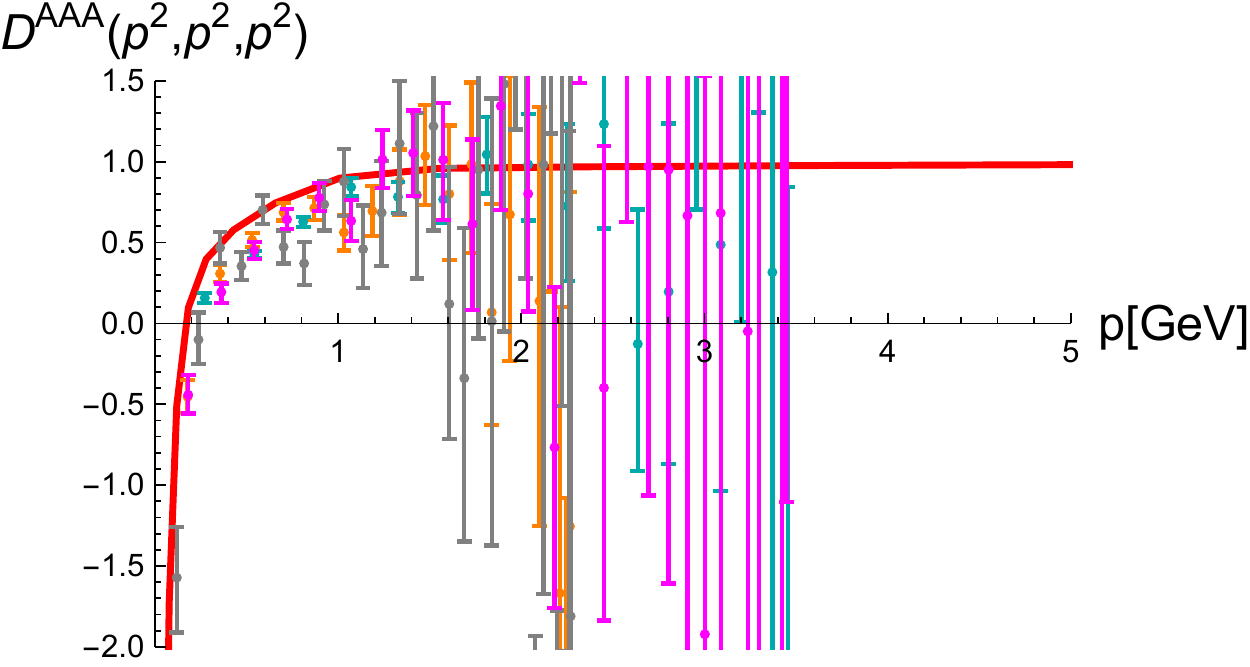}
\hfill
\includegraphics[width=0.47\textwidth]{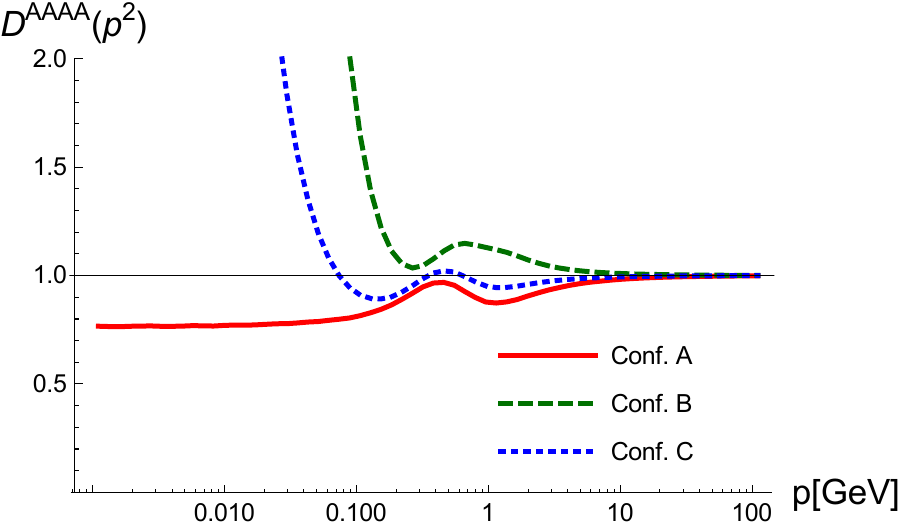}}
\caption{Left: Three-gluon vertex dressing from the full system in comparison with lattice results \cite{Cucchieri:2008qm}. Right: Tree-level dressing of the four-gluon vertex for different kinematic configurations.}
\label{fig:res-3g-4g}
\end{figure}

Some results for vertices are depicted in Fig.~\ref{fig:res-3g-4g}. The three-gluon vertex agrees very well with the lattice data. The ghost-gluon vertex, on the other hand, has a shifted maximum \cite{Huber:2016tvc}. The source of this deviation is currently unclear. For the four-gluon vertex three kinematic configurations are depicted in Fig.~\ref{fig:res-3g-4g}. The most notable feature is that deviations from the tree-level become only substantial at low momenta. In addition, a few selected dressing functions beyond the tree-level one were calculated which were found to be very small \cite{Huber:2016tvc}. It should be emphasized that this is not due to the smallness of single diagrams but stems from cancellations between diagrams. The same applies for the three-gluon vertex.

\begin{figure}[tb]
 \centerline{
 \includegraphics[width=0.47\textwidth]{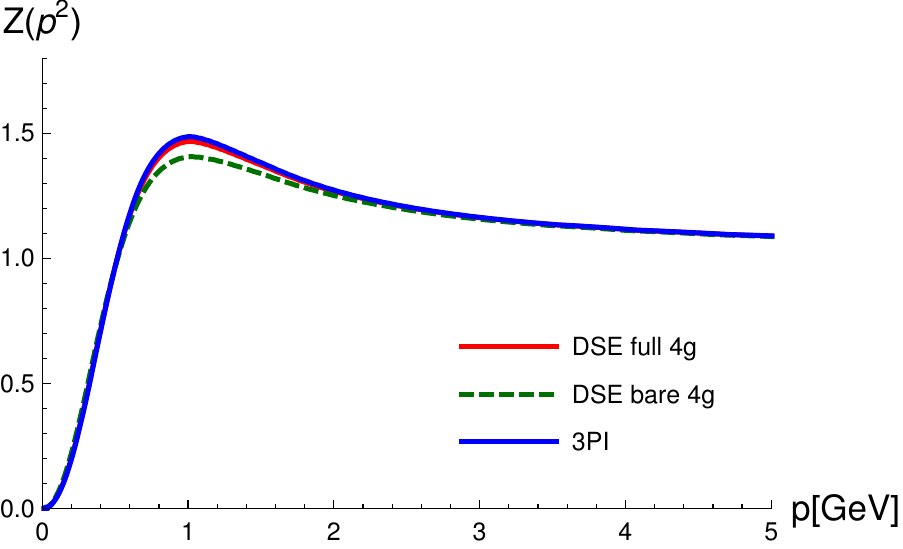}
 \hfill
 \includegraphics[width=0.47\textwidth]{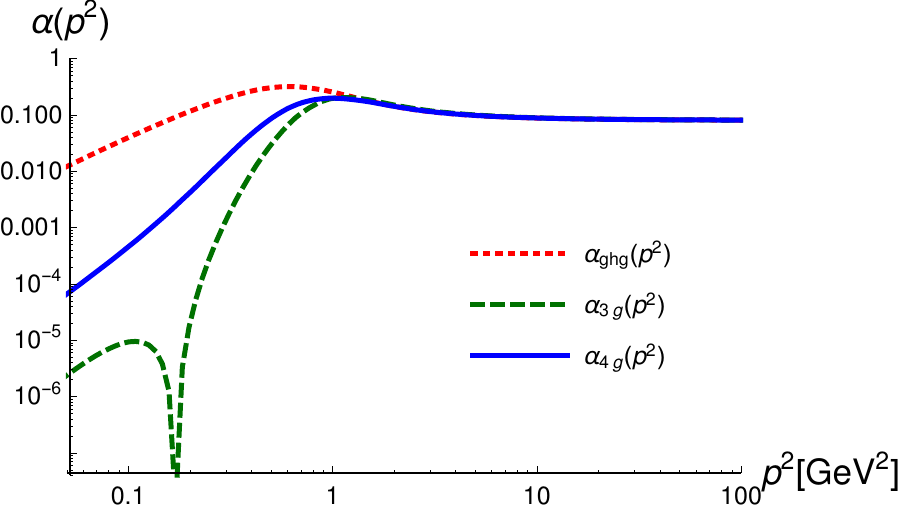}}
 \caption{Left: Gluon dressing function from three systems: DSEs with a dynamical or a bare four-gluon vertex and equations of motion from the 3PI effective action. Right: The coupling constants from the ghost-gluon, three-gluon and four-gluon vertices.}
 \label{fig:comp-coups}
\end{figure}

The stability of the truncation is tested by replacing the four-gluon vertex by a bare one. Another test consists in using the equations of motion of the $3$PI effective action. The results for the gluon dressing function are shown in Fig.~\ref{fig:comp-coups}. In the mid-momentum regime there is a small difference in the height of the bump. Larger differences are found in the deep IR, which become visible in the gluon propagator.

Fig.~\ref{fig:comp-coups} also shows the couplings as derived from the ghost-gluon, the three-gluon and the four-gluon vertices. It has to be noted that for this comparison not the MiniMOM coupling is used, but 
$$\alpha_\text{ghg}(p^2)=\frac{g^2}{4\pi} D^{A\bar c c}(p^2,p^2,p^2)G(p^2)^2Z(p^2)$$
which is also a suitable definition of a coupling \cite{Eichmann:2014xya}. The good agreement down to a few GeV and the hierarchy below coincides with the findings in four dimensions from the FRG in Ref.~\cite{Cyrol:2016tym}. The agreement between the different couplings in the perturbative regime shows that the Slavnov-Taylor identities are respected as discussed in \cite{Cyrol:2016tym}.

\section{Conclusions}

The results for the correlation functions of three-dimensional Yang-Mills theory proved to be stable against several modifications of the truncation. This might seem in contradiction to the deviation with available lattice results. However, it is currently not clear whether the obtained solution and the lattice solution should agree because of the possibility of several solutions. An important result is that the mild deviations of higher gluonic correlation functions from their tree-level behavior above $1\,\text{GeV}$ is caused by cancellations between diagrams and not by the smallness of single diagrams.

Given the many parallels between three- and four-dimensional Yang-Mills theory, it can be expected that these findings also apply in four dimensions. With the functional renormalization group using basically the same truncation as here (but producing a family of solutions) indeed good agreement with lattice results can be obtained \cite{Cyrol:2016tym}.

\section*{Acknowledgments}

Results have been obtained using the Vienna Scientific Cluster (VSC) and the HPC clusters at the University of Graz.
Funding by the FWF (Austrian science fund) under Contract No. P 27380-N27 is gratefully acknowledged.

\bibliographystyle{utphys_mod}
\bibliography{literature_YM_3d}

\end{document}